# Modelling the Behavior Classification of Social News Aggregations Users

Solomia Fedushko [*[0000-0001-7548-5856]], Olha Trach [[0000-0003-1461-791X]], Yaryna Turchyn [[0000-0002-9114-1911]], Zoryana Kunch [[0000-0002-8924-7274]], Ulyana Yarka [[0000-0003-1920-1980]]

Lviv Polytechnic National University, Lviv, Ukraine

```
solomiia.s.fedushko@lpnu.ua, olya@trach.com.ua,
turchyn.j@gmail.com, uliana.b.yarka@lpnu.ua,
zorjana.kunch@gmail.com
```

**Abstract.** This paper deals with actual fuzzy logic approach for modelling the behavior classification of social news aggregations users. The peculiarities of the structure of informational content of communities on the basis of social news aggregations are explored. A formal model of social news aggregation model has been developed, which includes user of the social news aggregation on the basis of fuzzy measures of its characteristics. The method of behavioral classification of users and methods for structuring sections and discussions of social news aggregations are developed. The methods for determining the main characteristics of the users of the social news aggregation: activeness, creativeness, attractiveness, reactiveness, loyalty, is developed. Method for defining characteristics and classification of social news aggregations users is presented.

## 1    Introduction

Each user of social news aggregations makes a contribution to the development of a social news aggregation. Contribution of the user can be determined objectively (based on the study of its behavior, information content, which is created and classification) and subjective (based on the assessments of other user of social news aggregations and expert evaluations). The indicator of user usefulness allows ranking the users of social news aggregations, using received information for further administrative measures. For example, the most useful users need to be involved in moderation of the social news aggregation, to stimulate the material, while users with negative usefulness need to remove from social news aggregations.
Ranking the users in the contribution and determining the core of social news aggregation user allows the administrator at any time to establish the users, who bring the social news aggregation the maximum benefit.

This information is necessary and critically useful for managing the social news aggregation, since the social news aggregation is different from the usual site, which the administrator must take into account the views and interests of the users. And in order to make social news aggregation-based solutions is needed to listen to users, who have a great authority and make the main contribution to its development.

The development of methods for behavioral classification of users of social news aggregation based on the presentation of information content as a tree is an urgent task.

## 2 The Method of User Behavioral Classification

The user of the social news aggregation a person who visits the social news aggregation site, reads or publishes its content in the form of discussions and messages on the social news aggregation is considered.

The model of the user of the social news aggregation is assigned in the following form:

$$SNAU_i = \langle Login_i, Password_i, Status_i, Email_i, MemName_i, LastVisit_i, PersonalData_i \rangle$$

where $SNAU_i$ is a user of social news aggregations; $Login_i$ is pseudo of user; $Password_i$ is password of user; $Status_i$ is the role of a user in the community; $Email_i$ is e-mail; $Memname_i$ is the name of the user; $LastVisit_i$ is date of the last visit to social news aggregation; $PersonalData_i$ is personal data of the user.

### 2.1 Development of Methods for the Calculating Characteristics of Social News Aggregation Users

In the course of the research it was established that the users of the social news aggregations (1) to a greater or lesser extent have the following characteristics in Figure 1.

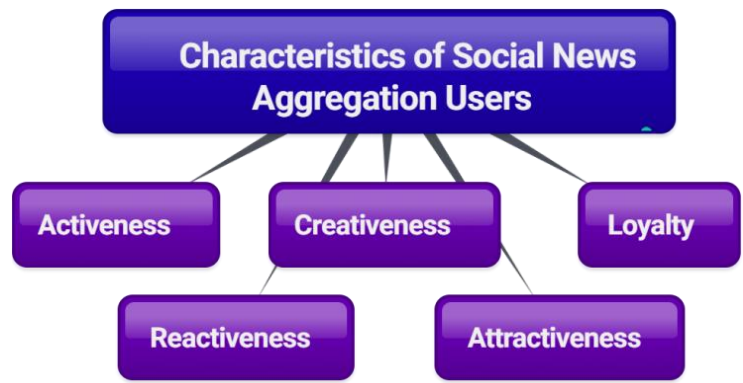

**Fig. 1.** Scheme of characteristics of social news aggregation users.

Based on these characteristics, we define the rules for classifying users in social news aggregations.

For presentation the characteristics of the users in the social news aggregation we will use a set of unclear plurals of the listed characteristics of the community users in relation to the entire social news aggregation are determined based on the analysis of the behavior of users within the community:

— activeness is determined by the amount of information content they create;
— creativeness is determined the quality of information content and how other users of community react to it;
— attractiveness is determined the quantity of users who react to the created content;
— reactiveness is a way of participating in discussions;
— loyalty is a reaction to the information content of other users.

### 2.2 Development of Methods for Calculating the Values of Linguistic Variables and Measures of Belonging

For each of the proposed characteristics of the users, we introduce the corresponding linguistic variables: Activeness, Creativeness, Attractiveness, Reactiveness, and Loyalty.

The linguistic variable is given by the quartet

$$< \beta, T, X, M >$$

where $\beta$ is the name of the linguistic variable;

$T$ is a plural of values of a linguistic variable, which is the names of fuzzy variables;

$$T = \{"low","medium","high"\}$$

$X$ is area of definition of fuzzy variables describing the linguistic variable

$$X = [0; card(Post)];$$

$M$ is a set of measures of fuzzy variables, which are values of a linguistic variable.

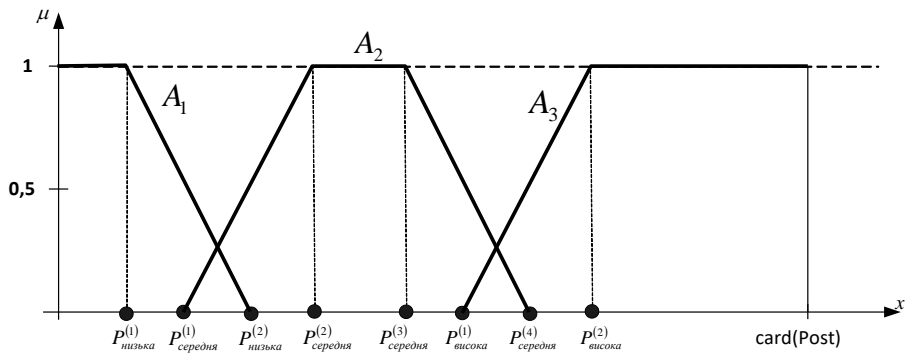

**Fig. 2.** The function of belonging to fuzzy plurals: $A_1$ –"low", $A_2$ – "medium", $A_3$ –"high".

$P_{low}^{(1)}, P_{low}^{(2)}, P_{medium}^{(1)}, P_{medium}^{(2)}, P_{medium}^{(3)}, P_{medium}^{(4)}, P_{high}^{(1)}, P_{high}^{(2)}$ are the parameters that are proportional to *card(Post)* and given by an expert or administrator of the forum, and moreover

$P_{low}^{1} \leq P_{medium}^{(1)} \leq P_{low}^{(2)} \leq P_{medium}^{(2)} \leq P_{medium}^{(3)} \leq P_{high}^{(1)} \leq P_{medium}^{(4)} \leq P_{high}^{(2)} < 1$.

All the P parameters that are used later in the membership functions are determined by experts for each feature of the social news aggregation.

We will write down the functions of membership for all characteristics (1) – (30) of the users of the social news aggregation.

**The activeness of creating discussions:**

$$\mu_{low}(Th) = \begin{cases} 1, \text{if } 0 \leq Activeness_{Thread}(SNAU_i) < Pat_{low}^{(1)} \\ \dfrac{Pat_{low}^{(2)} - Activeness_{Thread}(SNAU_i)}{Pat_{low}^{(2)} - Pat_{low}^{(1)}}, \\ \text{if } Pat_{low}^{(1)} \leq Activeness_{Thread}(SNAU_i) \leq Pat_{low}^{(2)} \end{cases} \quad (1)$$

$$\mu_{medium}(Th) = \begin{cases} \dfrac{Activeness_{Thread}(SNAU_i) - Pat_{medium}^{(1)}}{Pat_{medium}^{(2)} - Pat_{medium}^{(1)}}, \\ \text{if } Pat_{medium}^{(1)} \leq Activeness_{Thread}(SNAU_i) < Pat_{medium}^{(2)} \\ 1, \text{if } Pat_{medium}^{(2)} \leq Activeness_{Thread}(SNAU_i) \leq Pat_{medium}^{(3)} \\ \dfrac{Pat_{medium} - Activeness_{Thread}(SNAU_i)}{Pat_{medium}^{(3)} - Pat_{medium}^{(3)}}, \\ \text{if } Pat_{medium}^{(3)} < Activeness_{Thread}(SNAU_i) < Pat_{medium}^{(4)} \end{cases} \quad (2)$$

$$\mu_{high}(Th) = \begin{cases} \dfrac{Activeness_{Thread}(SNAU_i) - Pat_{high}^{(1)}}{Pat_{high}^{(2)} - Pat_{high}^{(1)}}, \\ \text{if } Pat_{high}^{(1)} \leq Activeness_{Thread}(SNAU_i) < Pat_{high}^{(2)} \\ 1, \text{if } Pat_{high}^{(2)} \leq Activeness_{Thread}(SNAU_i) \leq 1 \end{cases} \quad (3)$$

where $Activeness_{Thread}(SNAU_i)$ is activeness of creating posts j-th *user*.

**The activeness of creating polls:**

$$\mu_{low}(Pl) = \begin{cases} 1, & \text{if } 0 \leq Activeness_{Poll}(SNAU_i) \leq Papl_{low}^{(1)} \\ \dfrac{Papl_{low}^{(2)} - Activeness_{Poll}(SNAU_i)}{Papl_{low}^{(2)} - Papl_{low}^{(1)}}, \\ \text{if } Papl_{low}^{(1)} \leq Activeness_{Post}(SNAU_i) \leq Papl_{low}^{(2)} \end{cases} \quad (4)$$

$$\mu_{medium}(Pl) = \begin{cases} \dfrac{Activeness_{Poll}(SNAU_i) - Papl_{medium}^{(1)}}{Papl_{medium}^{(2)} - Papl_{medium}^{(1)}}, \\ \text{if } Papl_{medium}^{(1)} \leq Activeness_{Poll}(SNAU_i) < Papl_{medium}^{(2)} \\ 1, \text{ if } Papl_{medium}^{(2)} \leq Activeness_{Poll}(SNAU_i) < Papl_{medium}^{(3)} \\ \dfrac{Papl_{medium}^{(4)} - Activeness_{Poll}(SNAU_i)}{Papl_{medium}^{(3)} - Papl_{medium}^{(3)}}, \\ \text{if } Papl_{medium}^{(3)} \leq Activeness_{Poll}(SNAU_i) < Papl_{medium}^{(4)} \end{cases} \quad (5)$$

$$\mu_{high}(Pl) = \begin{cases} \dfrac{Activeness_{Poll}(SNAU_i) - Papl_{high}^{(1)}}{Papl_{high}^{(2)} - Papl_{high}^{(1)}}, \\ \text{if } Papl_{high}^{(1)} \leq Activeness_{Poll}(SNAU_i) \leq Papl_{high}^{(2)} \\ \\ 1, \text{ if } Papl_{high}^{(2)} \leq Activeness_{Poll}(SNAU_i) \leq 1 \end{cases} \quad (6)$$

where $Activeness_{Poll}(SNAU_i)$ is the activity of creating polls i-th user.

**The activeness of creating posts:**

$$\mu_{low}(Ps) = \begin{cases} 1, & \text{if } 0 \leq Activeness_{Post}(SNAU_i) \leq Paps_{low}^{(1)} \\ \dfrac{Paps_{low}^{(2)} - Activeness_{Post}(SNAU_i)}{Paps_{low}^{(2)} - Paps_{low}^{(1)}}, \\ \text{if } Paps_{low}^{(1)} \leq Activeness_{Post}(SNAU_i) \leq Paps_{low}^{(2)} \end{cases} \quad (7)$$

$$\mu_{medium}(Ps) = \begin{cases} \dfrac{Activeness_{Post}(SNAU_i) - Paps_{medium}^{(1)}}{Paps_{medium}^{(2)} - Paps_{medium}^{(1)}}, \\ \quad \text{if } Paps_{medium}^{(1)} \leq Activeness_{Post}(SNAU_i) < Paps_{medium}^{(2)} \\ 1, \text{ if } Paps_{medium}^{(2)} \leq Activeness_{Post}(SNAU_i) < Paps_{medium}^{(3)} \\ \dfrac{Paps_{medium}^{(4)} - Activeness_{Post}(SNAU_i)}{Paps_{medium}^{(3)} - Paps_{medium}^{(3)}}, \\ \quad \text{if } Paps_{medium}^{(3)} \leq Activeness_{Post}(SNAU_i) < Paps_{medium}^{(4)} \end{cases} \quad (8)$$

$$\mu_{high}(Ps) = \begin{cases} \dfrac{Activeness_{Post}(SNAU_i) - Paps_{high}^{(1)}}{Paps_{high}^{(2)} - Paps_{high}^{(1)}}, \\ \text{if } Paps_{high}^{(1)} \leq Activeness_{Post}(SNAU_i) \leq Paps_{high}^{(2)} \\ \\ 1, \text{ if } Paps_{high}^{(2)} \leq Activeness_{Post}(SNAU_i) \leq 1 \end{cases} \quad (9)$$

where $Activeness_{Post}(SNAU_i)$ is the activity of the *i*-th user in the creation of posts.

**Activeness of participation in polls:**

$$\mu_{low}(Vt) = \begin{cases} 1, \text{ if } 0 \leq Activeness_{Vote}(SNAU_i) \leq Pvt_{low}^{(1)} \\ \dfrac{Pvt_{low}^{(2)} - Activeness_{Vote}(SNAU_i)}{Pvt_{low}^{(2)} - P_{low}^{(1)}}, \\ \text{if } Pvt_{low}^{(1)} \leq Activeness_{Vote}(SNAU_i) \leq Pvt_{low}^{(2)} \end{cases} \quad (10)$$

$$\mu_{medium}(Vt) = \begin{cases} \dfrac{Activeness_{Vote}(SNAU_i) - Pvt_{medium}^{(1)}}{Pvt_{medium}^{(2)} - Pvt_{medium}^{(1)}}, \\ \quad \text{if } Pvt_{medium}^{(1)} \leq Activeness_{Vote}(SNAU_i) < Pvt_{medium}^{(2)} \\ 1, \text{ if } Pvt_{medium}^{(2)} \leq Activeness_{Vote}(SNAU_i) < Pvt_{medium}^{(3)} \\ \dfrac{Pvt_{medium}^{(4)} - Activeness_{Vote}(SNAU_i)}{Pvt_{medium}^{(3)} - Pvt_{medium}^{(3)}}, \\ \quad \text{if } Pvt_{medium}^{(3)} \leq Activeness_{Vote}(SNAU_i) < Pvt_{medium}^{(4)} \end{cases} \quad (11)$$

$$\mu_{high}(Vt) = \begin{cases} \dfrac{Activeness_{Vote}(SNAU_i) - Pvt_{high}^{(1)}}{Pvt_{high}^{(2)} - Pvt_{high}^{(1)}}, \\ \text{if } Pvt_{high}^{(1)} \leq Activeness_{Vote}(SNAU_i) \leq Pvt_{high}^{(2)} \\ \\ 1, \text{ if } Pvt_{high}^{(2)} \leq Activeness_{Vote}(SNAU_i) \leq 1 \end{cases} \quad (12)$$

where $Activeness_{Vote}(SNAU_i)$ is active participation in the voting.

**The activeness of evaluating the actions of other users:**

$$\mu_{low}(Fb) = \begin{cases} 1, \text{ if } 0 \leq Activity_{Feedback}(SNAU_i) < Pfb_{low}^{(1)} \\ \dfrac{Pfb_{low}^{(2)} - Activity_{Feedback}(SNAU_i)}{Pfb_{low}^{(2)} - Pfb_{low}^{(1)}}, \\ \text{if } Pfb_{low}^{(1)} \leq Activity_{Feedback}(SNAU_i) \leq Pfb_{low}^{(2)} \end{cases} \quad (13)$$

$$\mu_{medium}(Fb) = \begin{cases} \dfrac{Activeness_{Feedback}(SNAU_i) - Pfb_{medium}^{(1)}}{Pfb_{medium}^{(2)} - Pfb_{medium}^{(1)}}, \\ \text{if } Pfb_{medium}^{(1)} \leq Activeness_{Feedback}(SNAU_i) < Pfb_{medium}^{(2)} \\ 1, \text{ if } Pfb_{medium}^{(2)} \leq Activeness_{Feedback}(SNAU_i) \leq Pfb_{medium}^{(3)} \\ \dfrac{Pfb_{medium}^{(4)} - Activeness_{Feedback}(SNAU_i)}{Pfb_{medium}^{(3)} - Pfb_{medium}^{(3)}}, \\ \text{if } Pfb_{medium}^{(3)} < Activeness_{Feedback}(SNAU_i) < Pfb_{medium}^{(4)} \end{cases} \quad (14)$$

$$\mu_{high}(Fb) = \begin{cases} \dfrac{Activeness_{Feedback}(SNAU_i) - Pfb_{high}^{(1)}}{Pfb_{high}^{(2)} - Pfb_{high}^{(1)}}, \\ \text{if } P_{high}^{(1)} \leq Activeness_{Feedback}(meSNAU_imber_i) \leq Pfb_{high}^{(2)} \\ \\ 1, \text{ if } Pfb_{high}^{(2)} \leq Activeness_{Feedback}(SNAU_i) \leq 1 \end{cases} \quad (15)$$

where $Activeness_{Feedback}(SNAU_i)$ is the activeness of evaluating actions by the *i-th* user.

**Total activeness:**

Calculated based on these types of activeness:

$$\mu_{low}(Total) = \begin{cases} 1, & \text{якщо } 0 \leq Activeness_{Total}(SNAU_i) < Pt_{low}^{(1)} \\ \dfrac{Pt_{low}^{(2)} - Activeness_{Total}(SNAU_i)}{Pt_{low}^{(2)} - Pt_{low}^{(1)}}, \\ \text{if } Pt_{low}^{(1)} \leq Activeness_{Total}(SNAU_i) \leq Pt_{low}^{(2)} \end{cases} \quad (16)$$

$$\mu_{medium}(Total) = \begin{cases} \dfrac{Activeness_{Total}(SNAU_i) - Pt_{medium}^{(1)}}{Pt_{medium}^{(2)} - Pt_{medium}^{(1)}}, \\ \text{if } Pt_{medium}^{(1)} \leq Activeness_{Total}(SNAU_i) < Pt_{medium}^{(2)} \\ 1, \text{ if } Pat_{medium}^{(2)} \leq Activeness_{Total}(SNAU_i) \leq Pat_{medium}^{(3)} \\ \dfrac{Pat_{medium}^{(4)} - Activeness_{Thread}(SNAU_i)}{Pat_{medium}^{(3)} - Pat_{medium}^{(3)}}, \\ \text{if } Pat_{medium}^{(3)} < Activeness_{Total}(SNAU_i) < Pat_{medium}^{(4)} \end{cases} \quad (17)$$

$$\mu_{high}(Total) = \begin{cases} \dfrac{Activeness_{Total}(SNAU_i) - Pt_{high}^{(1)}}{Pt_{high}^{(2)} - Pt_{high}^{(1)}}, \\ \text{if } Pt_{high}^{(1)} \leq Activeness_{Total}(SNAU_i) < Pt_{high}^{(2)} \\ 1, \text{ if } Pt_{high}^{(2)} \leq Activeness_{Total}(SNAU_i) \leq 1 \end{cases} \quad (18)$$

where $Activeness_{Total}(SNAU_i)$ is total activeness of the i-th *user*.

**Creativeness of the user:**

$$\mu_{low}(Cr) = \begin{cases} 1, & \text{if } 0 \leq Creativeness(SNAU_i) < Pcr_{low}^{(1)} \\ \dfrac{Pcr_{low}^{(2)} - Creativeness(SNAU_i)}{Pcr_{low}^{(2)} - Pcr_{low}^{(1)}}, \\ \text{if } Pcr_{low}^{(1)} \leq Creativeness(SNAU_i) \leq Pcr_{low}^{(2)} \end{cases} \quad (19)$$

$$\mu_{medium}(Cr) = \begin{cases} \dfrac{Creativeness(SNAU_i) - Pcr_{medium}^{(1)}}{Pcr_{medium}^{(2)} - Pcr_{medium}^{(1)}}, \\ \quad \text{if } Pcr_{medium}^{(1)} \leq Creativeness(SNAU_i) < Pcr_{medium}^{(2)} \\ 1, \quad \text{if } Pcr_{medium}^{(2)} \leq Creativeness(SNAU_i) \leq Pcr_{medium}^{(3)} \\ \dfrac{Pcr_{medium}^{(4)} - Creativeness(SNAU_i)}{Pcr_{medium}^{(3)} - Pcr_{medium}^{(3)}}, \\ \quad Pcr_{medium}^{(3)} < Creativeness(SNAU_i) < Pcr_{medium}^{(4)} \end{cases} \quad (20)$$

$$\mu_{high}(Cr) = \begin{cases} \dfrac{Creativeness(SNAU_i) - Pcr_{high}^{(1)}}{Pcr_{high}^{(2)} - Pcr_{high}^{(1)}}, \\ \text{if } Pcr_{high}^{(1)} \leq Creativeness(SNAU_i) < Pcr_{high}^{(2)} \\ \\ 1, \quad \text{if } Pcr_{high}^{(2)} \leq Creativeness(SNAU_i) \leq 1 \end{cases} \quad (21)$$

where $Creativeness(SNAU_i)$ is creativeness of the user.

**Atractiveness of the user:**

$$\mu_{low}(Attr) = \begin{cases} 1, \quad \text{if } 0 \leq Attractiveness(SNAU_i) < Pattr_{low}^{(1)} \\ 1\dfrac{Pattr_{low}^{(2)} - Attractiveness(SNAU_i)}{Pattr_{low}^{(2)} - Pattr_{low}^{(1)}}, \\ \text{if } Pattr_{low}^{(1)} \leq Attractiveness(SNAU_i) \leq Pattr_{low}^{(2)} \end{cases} \quad (22)$$

$$\mu_{medium}(Attr) = \begin{cases} \dfrac{Attractiveness(SNAU_i) - Pattr_{medium}^{(1)}}{Pattr_{medium}^{(2)} - Pattr_{medium}^{(1)}}, \\ \quad \text{if } Pattr_{medium}^{(1)} \leq Attractiveness(SNAU_i) < Pattr_{medium}^{(2)} \\ 11, \quad \text{if } Pattr_{medium}^{(2)} \leq Attractiveness(SNAU_i) \leq Pattr_{medium}^{(3)} \\ \dfrac{Pattr_{medium}^{(4)} - Attractiveness(SNAU_i)}{Pattr_{medium}^{(3)} - Pattr_{medium}^{(3)}}, \\ \quad \text{if } Pattr_{medium}^{(3)} < Attractiveness(SNAU_i) < Pattr_{medium}^{(4)} \end{cases} \quad (23)$$

$$\mu_{high}(Attr) = \begin{cases} \dfrac{Attractiveness(SNAU_i) - Pattr_{high}^{(1)}}{Pattr_{high}^{(2)} - Pattr_{high}^{(1)}}, \\ 1 \text{ if } Pattr_{high}^{(1)} \leq Attractiveness(SNAU_i) < Pattr_{high}^{(2)} \\ \\ 1, \text{ if } Pattr_{high}^{(2)} \leq Attractiveness(SNAU_i) \leq 1 \end{cases} \quad (24)$$

where $Attractiveness(SNAU_i)$ is attractiveness of the user.

**Reactiveness of the user.:**

$$\mu_{low}(R) = \begin{cases} 1, \text{ if } 0 \leq Reactiveness(SNAU_i) < Pr_{low}^{(1)} \\ 1 \dfrac{Pr_{low}^{(2)} - Reactiveness(SNAU_i)}{Pr_{low}^{(2)} - Pr_{low}^{(1)}}, \\ \text{ if } Pr_{low}^{(1)} \leq Reactiveness(SNAU_i) \leq Pr_{low}^{(2)} \end{cases} \quad (25)$$

$$\mu_{medium}(R) = \begin{cases} \dfrac{Reactiveness(SNAU_i) - Pr_{medium}^{(1)}}{Pr_{medium}^{(2)} - Pr_{medium}^{(1)}}, \\ \text{ if } Pr_{medium}^{(1)} \leq Reactiveness(SNAU_i) < Pr_{medium}^{(2)} \\ 1, \text{ if } Pr_{medium}^{(2)} \leq Reactiveness(SNAU_i) \leq Pr_{medium}^{(3)} \\ \dfrac{Pr_{medium}^{(4)} - Reactiveness(SNAU_i)}{Pr_{medium}^{(3)} - Pr_{medium}^{(3)}}, \\ \text{ if } Pr_{medium}^{(3)} < Reactiveness(SNAU_i) < Pr_{medium}^{(4)} \end{cases} \quad (26)$$

$$\mu_{high}(R) = \begin{cases} \dfrac{Reactiveness(SNAU_i) - Pr_{high}^{(1)}}{Pr_{high}^{(2)} - Pr_{high}^{(1)}}, \\ \text{ if } Pr_{high}^{(1)} \leq Reactiveness(SNAU_i) < Pr_{high}^{(2)} \\ \\ 1, \text{ if } Pr_{high}^{(2)} \leq Reactiveness(SNAU_i) \leq 1 \end{cases} \quad (27)$$

where $Reactiveness(SNAU_i)$ is reactiveness of the user.

**Loyalty of the user:**

$$\mu_{low}(L) = \begin{cases} 1, & \text{if } 0 \leq Loyalty(SNAU_i) < Pl_{low}^{(1)} \\ \dfrac{Pl_{low}^{(2)} - Loyalty(SNAU_i)}{Pl_{low}^{(2)} - Pl_{low}^{(1)}}, \\ \text{if } Pl_{low}^{(1)} \leq Loyalty(SNAU_i) \leq Pl_{low}^{(2)} \end{cases} \qquad (28)$$

$$\mu_{medium}(L) = \begin{cases} \dfrac{Loyalty(SNAU_i) - Pl_{medium}^{(1)}}{Pl_{medium}^{(2)} - Pl_{medium}^{(1)}}, \\ \text{if } Pl_{medium}^{(1)} \leq Loyalty(SNAU_i) < Pl_{medium}^{(2)} \\ 1, & \text{if } Pl_{medium}^{(2)} \leq Loyalty(SNAU_i) \leq Pl_{medium}^{(3)} \\ \dfrac{Pl_{medium}^{(4)} - Loyalty(SNAU_i)}{Pl_{medium}^{(3)} - Pl_{medium}^{(3)}}, \\ \text{if } Pl_{medium}^{(3)} < Loyalty(SNAU_i) < Pl_{medium}^{(4)} \end{cases} \qquad (29)$$

$$\mu_{high}(L) = \begin{cases} \dfrac{Loyalty(SNAU_i) - Pl_{high}^{(1)}}{Pl_{high}^{(2)} - Pl_{high}^{(1)}}, \\ \text{if } Pl_{high}^{(1)} \leq Loyalty(SNAU_i) < Pl_{high}^{(2)} \\ 1, & \text{if } Pl_{high}^{(2)} \leq Loyalty(SNAU_i) \leq 1 \end{cases} \qquad (30)$$

where $Loyalty(SNAU_i)$ is loyalty of the user.

## 3   Building Rules for Classifying Users of Social News Aggregation

The classification rules for each class of users of the social news aggregation are formulated based on the developed methods for calculating the characteristics of users and certain classes of social news aggregation users. The classes of the social news aggregation are proposed:

- Activist
- Moderator
- Flamer
- Author
- Critic
- Reader.

The membership of the users in one of the classes based on its characteristics (Activity, Creativity, Attraction, Reactivity, Loyalty) is represented by production rules and Table 1.

**Table 1.** Classes users of the social news aggregation.

|                | Moderator        | Flamer           | Author           | Critic           | Reader           | Activist      |
|----------------|------------------|------------------|------------------|------------------|------------------|---------------|
| Activeness     | medium, high     | medium, high     | medium, high     | low              |                  | low           |
| Creativeness   | medium, high     |                  | low              | medium, high     | low              | low           |
| Attractiveness |                  |                  |                  | high             | low, medium      | low, medium   |
| Reactiveness   | low, medium      | high             |                  | low              | medium, high     | low           |
| Loyalty        |                  | medium, high     | low              |                  |                  |               |

1. If $Activeness(SNAU) \in \{"medium","high"\}$ and $Creativeness(SNAU) \in \{"medium","high"\}$ and $Reactivenes(SNAU) \in \{"low","medium"\}$ then $Member - Activist;$

2. If $Activeness(SNAU) \in \{"medium","high"\}$ and $Reactivenes(SNAU) = "high"$ and $Loyatly(SNAU) \in \{"medium","high"\}$, then $Member - Moderator;$

3. If $Activeness(SNAU) \in \{"medium","high"\}$ and $Creativeness(SNAU) = "low"$ and $Loyatly(SNAU) = "low"$, then $Member - Babbler;$

4. If $Activeness(SNAU) = "low"$ and $Creativeness(SNAU) \in \{"medium","high"\}$ and $Atractiveness(SNAU) \in \{"medium","high"\}$ and $Reactiveness(SNAU) = "low"$, then $Member - Author;$

5. If $Creativeness(SNAU) = "low"$ and $Reactiveness(SNAU) \in \{"low","high"\}$ and $Atractiveness(SNAU) = "high"$ and $Loyatly(SNAU) = "low"$, then $Member - Critic;$

6. If $Atractiveness(SNAU) = "low"$ and $Creativeness(SNAU) = "low"$ and $Atractiveness(SNAU) \in \{"low","average"\}$ and $Reactiveness(SNAU) = "low"$, then $Member - Reader;$

## 4    Determine the User's Usefulness for the Community

The usefulness of a user for the social news aggregation is a complex indicator calculated on the basis of its characteristics: activeness, attractiveness, creativeness, reactiveness and loyalty. The usefulness of user *ME* is calculated by the equation:

$$ME = C_1 \times Activeness + C_2 \times Attractiveness + C_3 \times Creativeness + \\ + C_4 \times Reactiveness + C_5 \times Loyalty \quad (31)$$

where $C_1, C_2, \ldots, C_5$ is weight coefficients of each user's characteristics, which are determined based on the development scenario of the social news aggregation, moreover $\sum_i C_i = 1$, $C_i \geq 0$.

Consequently, $ME \in [0, 1]$.

The user's usefulness allows the administrator to evaluate the importance of the user for the community and to take this value into account when applying sanctions.

## 5    Conclusion

In this work the models have been developed that are the basis for further research on the construction of effective site positioning methods. Formalized structure of the social news aggregation, which includes two components (information content, users) is suggested.

The peculiarities of the structure of informational content of communities on the basis of social news aggregations are explored. A formal social news aggregation model has been developed, which includes the model of user of the social news aggregation on the basis of fuzzy measures of its characteristics, the model of the structure of information content and the model of the content of information content, on the basis of which developed the method of behavioral classification of users and methods for structuring sections and discussions of social news aggregations.

The methods for determining the main characteristics of the users of the web community: activeness, creativeness, attractiveness, reactiveness, loyalty are developed.

The classes of users of social news aggregations on the basis of social news aggregations are allocated and the rules of classification of users are formulated.

## References

1. Jin, L., Chen, Y., Wang, T., Hui, P., Vasilakos, A.: Understanding user behavior in online social networks: a survey. Communications Magazine, 51, (9), 144-150 (2013).
2. Benevenuto, F., Rodrigues, T., Cha, M., Almeida, V.: Characterizing user behavior in online social networks. In: 9th ACM SIGCOMM conference on Internet measurement (IMC '09). ACM, pp. 49-62. New York, USA (2009).

3. Fernández, A., García, S., del Jesus, M. J., Herrera, F.: A study of the behaviour of linguistic fuzzy rule based classification systems in the framework of imbalanced data-sets. Fuzzy Sets Syst., 159 (18), 2378-2398 (2008).
4. Korzh, R., Peleshchyshyn, A., Syerov, Y., Fedushko, S.: University's information image as a result of university web communities' activities. Advances in Intelligent Systems and Computing. Advances in Intelligent Systems and Computing, 512, pp. 115-127. Springer, Cham (2017). DOI: 10.1007/978-3-319-45991-2_8
5. Morente-Molinera, J., et al.: Supervised learning classification methods using multigranular linguistic modeling and fuzzy entropy, Transactions on Fuzzy Systems, 25(5), 1078-1089 (2017).
6. Fedushko, S., Ustyianovych, T.: Predicting pupil's successfulness factors using machine learning algorithms and mathematical modelling methods. Advances in Computer Science for Engineering and Education II. ICCSEEA 2019. Advances in Intelligent Systems and Computing, 938, pp. 625-636. Springer (2020). DOI 10.1007/978-3-030-16621-2_58
7. Olson, D. L., Delen, D: Advanced Data Mining Techniques. NY, USA: Springer, 2008.
8. Lee, H.-M., et al.: An efficient fuzzy classifier with feature selection based on fuzzy entropy, Trans. Syst. Man Cybern. B Cybern., 31 (3), 426-432 (2001).
9. Wendy, N., Misha, P.: Moving behavioral theories into the 21st century: technological advancements for improving quality of life. IEEE pulse. 4(5), 25-28 (2013).
10. Syerov, Y., Fedushko, S., Loboda, Z.: Determination of development scenarios of the educational web forum. In: 11th International Scientific and Technical Conference "Computer Sciences and Information Technologies" CSIT-2016, 73-76 (2016). DOI: 10.1109/STC-CSIT.2016.7589872.
11. Liu, J., Weitzman, E., Chunara, R.: Assessing Behavioral Stages From Social Media Data. In: Conference on Computer-Supported Cooperative Work, pp.1320-1333 (2017).
12. Bilushchak, T., Peleshchyshyn, A., Komova, M.: Development of method of search and identification of historical information in the social environment of the Internet. In: XIth International Scientific and Technical Conference "Computer Sciences and Information Technologies" CSIT-2017, 196-199 (2017).
13. Trach O., Peleshchyshyn A.: Functional-network model of tasks performance of virtual communication life cycle directions. In: XIIth International Conference "Computer Sciences and Information Technologies" CSIT-2016. pp. 108-110. Lviv (2016).
14. Trach O., Vus V., Tymovchak-Maksymets O.: Typical algorithm of stage completion when creating a virtual community of a HEI. In: XIIIth International Conference on Advanced Trends in Radioelectronics, Telecommunications and Computer Engineering, TCSET-2016, pp. 849-851. Lviv-Slavske (2016).
15. Dosyn, D., Lytvyn, V., Kovalevych, V., Oborska, O., Holoshchuk, R.: Knowledge discovery as planning development in knowledgebase framework. Modern Problems of Radio Engineering, Telecommunications and Computer Science. In: 13th International Conference on on Advanced Trends in Radioelectronics, Telecommunications and Computer Engineering, TCSET-2016, pp. 449-451. Lviv-Slavsko, Ukraine (2016).
16. Lytvyn, V., Peleshchak, I., Peleshchak, R., Holoshchuk, R.: Detection of multispectral input images using nonlinear artificial neural networks. In: 14th International Conference on Advanced Trends in Radioelectronics, Telecommunications and Computer Engineering, TCSET, pp. 119-122. Slavske, Ukraine (2018).
17. Artem, K., Holoshchuk, R., Kunanets, N., Shestakevysh, T., Rzheuskyi, A.: Information Support of Scientific Researches of Virtual Communities on the Platform of Cloud Services. Advances in Intelligent Systems and Computing. Advances in intelligent systems and computing III, 871, pp. 301-311 (2019).

18. Zhezhnych, P., Tarasov, D.: Methods of data processing restriction in ERP Systems. In: 13th International Scientific and Technical Conference on Computer Sciences and Information Technologies, CSIT 2018, pp. 274-277 (2018).
19. Tkachenko, R., Izonin, I.: Model and Principles for the Implementation of Neural-Like Structures based on Geometric Data Transformations. Advances in Computer Science for Engineering and Education. International Conference on Computer Science, Engineering and Education Applications, ICCSEEA 2018. Advances in Intelligent Systems and Computing, vol 754, 578-587. Springer, Cham (2019). doi: 10.1007/978-3-319-91008-6_58
20. Gozhyj, A., Vysotska, V., Yevseyeva, I., Kalinina, I., Gozhyj, V.: Web resources management method based on intelligent technologies. Advances in Intelligent Systems and Computing, 871, 206-221 (2019).
21. Gozhyj, A., Kalinina, I., Vysotska, V., Gozhyj, V.: The method of web-resources management under conditions of uncertainty based on fuzzy logic. In: 13th International Scientific and Technical Conference "Computer Sciences and Information Technologies" CSIT 2018, pp. 343-346 (2018).
22. Lytvyn, V., Dosyn, D., Emmerich, M., Yevseyeva, I.: Content formation method in the web systems. CEUR Workshop Proceedings, 2136, pp. 42-61 (2018).
23. Anisimova O., Vasylenko V.: Social networks as the instrument for a higher education institution image creation. 1-st International Workshop Control, Optimization and Analytical Processing of Social Networks (2019). In press.